\documentstyle[epsfig]{mn}

\begin{document}



\def\be{\begin{equation}}
\def\ee{\end{equation}}
\def\mdot{$\dot{m}$ }
\def\mpc{\,{\rm {Mpc}}}
\def\kpc{\,{\rm {kpc}}}
\def\kms{\,{\rm {km\, s^{-1}}}}
\def\msun{{$M_{\odot}$}}
\def\Gyr{{\,\rm Gyr}}
\def\erg{{\rm erg}}
\def\sr{{\rm sr}}
\def\hz{{\rm Hz}}
\def\cm{{\rm cm}}
\def\sec{{\rm s}}
\def\eV{{\rm \ eV}}
\def\ledd{$L_{Edd}$~}
\def\mic{$\mu$ }
\def\ang{\AA }  
\def\cm2{cm$^2$ }
\def\se1{s$^{-1}$ }

\def\arcmin{\hbox{$^\prime$} }
\def\arcsec{\hbox{$^{\prime\prime}$} }
\def\degree{$^{\circ}$ } 
\def\mic{$\mu$ }
\def\ang{\AA }  
\def\cm2{cm$^2$ }
\def\se1{s$^{-1}$ }

\def\gtsima{$\; \buildrel > \over \sim \;$}
\def\ltsima{$\; \buildrel < \over \sim \;$}
\def\prosima{$\; \buildrel \propto \over \sim \;$}
\def\gsim{\lower.5ex\hbox{\gtsima}}
\def\lsim{\lower.5ex\hbox{\ltsima}}
\def\simgt{\lower.5ex\hbox{\gtsima}}
\def\simlt{\lower.5ex\hbox{\ltsima}}
\def\simpr{\lower.5ex\hbox{\prosima}}
\def\la{\lsim}
\def\ga{\gsim}
\def\Lsun{\L_\odot}
\def\sr{4U~1708--40}

\def\ie{{\frenchspacing\it i.e. }}
\def\eg{{\frenchspacing\it e.g. }}
\def\etal{{\it et al.~}}

\title[Radio:X--ray correlation in low/hard state BHCs]  
{A universal radio:X--ray correlation in low/hard state black hole binaries} 
\author[E. Gallo, R. P. Fender \& G. G. Pooley]  
{E. Gallo$^{1}$\thanks{egallo@science.uva.nl},
R. P. Fender$^{1}$, 
G. G. Pooley$^{2}$ 
\\ \\
$^{1}$ Astronomical Institute `Anton Pannekoek' and Center for High Energy
Astrophysics, University of Amsterdam, Kruislaan 403, \\
1098 SJ Amsterdam, The Netherlands. \\
$^{2}$ Mullard Radio Astronomy Observatory, Cavendish Laboratory, Madingley
Road, Cambridge CB3 0HE, UK.\\} 
\maketitle
\begin{abstract} 
Several independent lines of evidence now point to a connection 
between the physical processes that govern radio (\ie jet) and X--ray
emission from accreting X--ray binaries.    
We present a comprehensive study of (quasi--)simultaneous
radio:X--ray observations of stellar black hole binaries during the
spectrally hard X--ray state, finding evidence for a strong correlation between
these two bands over more than three orders of magnitude in X--ray luminosity.
The correlation extends 
from the quiescent regime up to close to the soft state transition, where radio 
emission starts to decline, sometimes below detectable levels, probably
corresponding to the physical disappearance of the jet. 
The X--ray transient V 404 Cygni is found to display the same functional
relationship already reported for GX 339--4 between radio and X--ray flux,
namely $S_{radio}\propto S_{X}^{+0.7}$.
In fact the data for all low/hard state black holes is consistent with a
universal relation between the radio and X--ray luminosity of the form 
$L_{radio}\propto L_{X}^{+0.7}$. 
Under the hypothesis of common physics
driving the disc--jet coupling in different sources, 
the observed spread   
to the best--fit relation can be interpreted in terms of a distribution in
Doppler factors and hence used to constrain the bulk Lorentz factors of both the
radio and X--ray emitting regions.  
Monte Carlo simulations show
that, assuming little or no X--ray beaming, the measured scatter in radio
power is consistent with Lorentz factors $\simlt2$ for the outflows in the
low/hard state, significantly less relativistic than the jets associated with
X--ray transients.    
When combined radio and X--ray beaming is considered, the range of possible
jet bulk velocities significantly broadens, allowing highly relativistic
outflows, but implying therefore severe X--ray selection effects. 
If the radio luminosity scales as the total jet power raised to $x>0.7$, 
then there exists an X--ray luminosity below which most of   
the accretion power will be channelled into the jet, rather than in the
X--rays. For $x=1.4$, as in several optically thick jet models, the
power output of `quiescent' black holes may be jet--dominated below $L_X\simeq
4 \times 10^{-5}$ \ledd. 
\end{abstract}  
\section{Introduction}
There is strong observational evidence for the fact that 
powerful radio--emitting outflows form a key part of the accretion
behaviour in some states of X--ray binary systems.
Due to its high brightness temperature, `nonthermal' spectrum and, in some
cases, high degree of polarisation, radio emission from black hole binaries is
believed to originate in 
synchrotron radiation from relativistic electrons ejected by the system with
large bulk velocities (Hjellming \& Han 1995; Mirabel \& Rodr\'\i guez 1999;
Fender, 2000, 2001a,b,c).  \\ 
Black hole binary systems are traditionally classified by their X--ray
features (see Nowak 1995; Poutanen 1998; Done 2001; Merloni
2002 for recent reviews), namely: a) the relative strength of a soft `black
body' component around 1 keV, b) the 
spectral hardness at higher energies c) X--ray luminosity and 
d) timing properties. Different radio behaviour is associated with
several `X--ray states', according to the following broad scheme.  
The \emph{low/hard state} is dominated by a power--law spectrum,
with a relatively low luminosity and an exponential cut--off
above about 100 keV and little or no evidence for a soft,
thermal component.
It is associated with a steady, self--absorbed
outflow that emits synchrotron radiation in the radio 
(and probably infrared) spectrum.
The \emph{quiescent/off state}, characterised by an extremely low X--ray flux,
may simply be interpreted as the hard state `turned down'    
to lower accretion rates and radiative efficiency.   
X--ray spectra from \emph{high/soft state} 
Black Hole Candidates (BHCs) are instead dominated by thermal radiation, while
the core radio emission drops 
below detectable levels, probably corresponding to the physical suppression
of the jet. In the \emph{very high state} both the thermal and the
power law components contribute substantially to the spectral energy
distribution. At a lower luminosity level an \emph{intermediate state} is also
observed, with     
properties similar to those of the very high state. For both the very high and
the intermediate state the connection with radio behaviour is not yet
clearly established. Corbel \etal (2001) show that the radio emission
from XTE 1550--564 in the intermediate state was suppressed by a factor $>50$
with respect to the hard    
state, while Homan \etal (2001) claim that intermediate and very high states can actually
occur at a wide range of luminosities.  \\ 
Transitions between   
states are often associated with multiple ejections of synchrotron emitting
material, possibly with high bulk Lorentz factors (Hjellming \& Han 1995;
Kuulkers \etal 1999; Fender \& Kuulkers 2001).\\ 
As already mentioned, BHCs in the low/hard state,  
like Cygnus X--1 and GX 339--4, are characterised by a 
flat or slightly inverted radio spectrum  ($\alpha = 
\Delta \log S_{\nu} / \Delta \log \nu  \simeq 0$), interpreted as arising from
a collimated, self--absorbed compact jet, in analogy to those observed in
active galactic nuclei (Blandford \& K\"onigl 1979). With the direct imaging
of a resolved compact radio jet from   
the core of Cygnus X--1 (Stirling \etal 2001), this association has been
confirmed. Radio emission from X--ray binaries, especially the
BHCs, is increasingly interpreted as the  
radiative signature of jet--like outflows. \\
It has been generally accepted that the soft thermal component of BH spectra
originates in an optically thick, geometrically thin accretion disc (Shakura
\& Sunyaev 1973), whereas the power law component      
is produced by Comptonisation of `seed' photons in a hot, rarefied `corona' of 
(quasi--) thermal electrons (Shapiro, Lightman \& Eardley 1976; Sunyaev \&
Titarchuk 1980; Haardt \& Maraschi 1991; Poutanen \& Svensson 1996). 
Although this picture can successfully reproduce the X--ray behaviour, it can
not yet address the clear correlation between radio and X--ray 
emission established for the persistent BHCs GX 339--4 and Cygnus X--1 
while in the hard state (Hannikainen \etal 1998; Brocksopp \etal 1999; Corbel
\etal 2000; Corbel \etal 2003). Moreover, some hard state sources, like XTE
J1118+480 and GX 339--4, show evidence for a 
turnover in the infrared--optical band, where the flat--to--inverted radio
spectrum seems to connect to an optically thin component extending up to the
X--rays (Corbel \& Fender 2002; Markoff \etal 2003a,b and references therein),
suggesting again that the jet plays a role at higher frequencies. \\
\\ 
Hence, all the evidence points to the corona in these systems being physically
related to the presence of a jet: by far the simplest interpretation therefore
is that the Comptonising region is just the base of the relativistic outflow
(Fender \etal 1999b; Merloni \& Fabian 2002; Markoff \etal 2003a).  
However, joining these two previously independent scenarios is somewhat
problematic because they often require different electron distributions and
geometries.\\ 
Due to the fast timescales in X--ray binary systems,
only simultaneous radio and X--ray observations provide the necessary tools to
probe this conjecture.  
The following results extend and complete those presented in
Gallo, Fender \& Pooley (2002).  

\begin{table*}
\label{tab10bhs}
\caption{System parameters for the ten hard state BHCs under consideration. 
Distance and $N_{H}$ references are   
given in parentheses next to each value. Inclination and BH mass estimates,
unless differently specified, 
are all taken from Orosz (2002). 
The last column refers to the literatures' sources from which we have obtained
(quasi--)
simultaneous radio and X--ray fluxes; no reference appears in case of our
own observations (Cygnus X--1). } 
\centering
\begin{tabular}{rccccl}
\hline
\hline
Source           	  & Distance(ref)  &Inclination(12)    	&BH mass(12) 	&N$_{H}$(ref)	& Data \\
                		  & (kpc)          & (degree)		&(\msun) 		&(10$^{21}$cm$^{-2}$)&\\ 
\hline
Cygnus X--1      	&  2.1 (1)       & 35$\pm$5          	&6.85--13.25   	&6.2  (16)  	& --\\
V 404 Cygni         	&  3.5 (2)       & 56$\pm$4          	&10.06--13.38  	&5.0 (17)  	&21,22\\
GRS 1758--258    	&  8.5 (3)       &   ?&$\sim$8--9 (14)	&14.0 (3)   	& 23\\
XTE J1118+480    	&  1.8 (4)       & 81$\pm$2          	&6.48--7.19	&0.1 (18)   	&24,25\\ 
GRO J0422+32     	&  2.4 (5)       & 44$\pm$2          	&3.66--4.97    	&2.0 (5)   	&24\\
GX 339--4        	&  4.0 (6)       & 15--60 (13)       	&5.8$\pm$0.5 (15)&6.0 (6)   	& 26\\
1E 1740.7--2942  	&  8.5 (7)       &   ?            	&    ?         	&118 (19) 	&27\\
XTE J1550--564   	&  4.0 (8,9$^{*}$) & 72$\pm$5          	&8.36--10.76	&8.5 (20)  	&20,28\\
GS 1354--64      	&  10.0 (10)      &   ?               		&     ?        		&32.0 (9)   	&29\\
4U 1543--47      	&  9.0 (11)      & 20.7$\pm$1.5      	&8.45--10.39   	&3.5  (11)  	&30\\
\hline
\hline
\end{tabular}
\flushleft
{\bf References :} 
\bf 1:  \rm Massey \etal 1995;
\bf 2:  \rm Zycki, Done \& Smith 1999;
\bf 3:  \rm Main \etal 1999;
\bf 4:  \rm McClintock \etal 2001;
\bf 5:  \rm Shrader \etal 1997;
\bf 6:  \rm Zdziarski \etal 1998;
\bf 7:  \rm Sunyaev \etal 1991;  
\bf 8:  \rm Kong \etal 2002;
\bf 9:  \rm Tomsick \etal 2001;
\bf 10:  \rm Kitamoto \etal 1990;
\bf 11: \rm Orosz \etal 1998;
\bf 12: \rm Orosz 2002;
\bf 13: \rm Cowley \etal 2002;
\bf 14: \rm Keck \etal 2001;
\bf 15: \rm Hynes \etal 2003;
\bf 16: \rm Schulz \etal 2002;
\bf 17: \rm Wagner \etal 1994;
\bf 18: \rm Dubus \etal 2001;
\bf 19: \rm Gallo \& Fender 2002;
\bf 20: \rm Tomsick \etal 2001;
\bf 21: \rm Han \& Hjellming 1992;
\bf 22: \rm Hjellming \etal 2000;
\bf 23: \rm Lin \etal 2000;
\bf 24: \rm Brocksopp \etal 2003;
\bf 25: \rm Markoff, Falkce \& Fender 2001;
\bf 26: \rm Corbel \etal 2000;
\bf 27: \rm Heindl, Prince \& Grunsfeld 1994;
\bf 28: \rm Corbel \etal 2001;
\bf 29: \rm Brocksopp \etal 2001;
\bf 30: \rm Brocksopp, private communication.\\
\bf $^{*}$\rm For XTE J1550--564 a distance of 4 kpc is assumed by both 
Kong \etal (2002) and Tomsick \etal (2001), as average value between 2.5 and 6
kpc, given by S\'anchez--Fern\'andez \etal (1999) and Sobczak \etal (1999)
respectively. 
\end{table*}
\section{The sample} 
\begin{figure}
\hspace{-0.7cm}
\psfig{figure=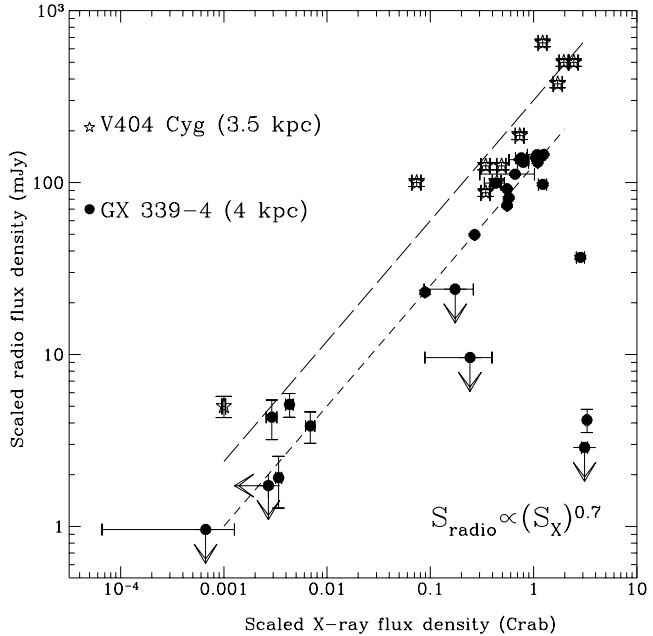,height=9cm}
\caption{\label{fit}Radio against X--ray flux density,  
scaled to a distance of 1 kpc and absorption corrected, 
for V 404 Cygni and GX 339--4.
Lines denote the fits to the datasets: short and long dashed for GX
339--4 and V 404 Cygni respectively. It is found that the    
data of V 404 Cygni are well fitted by the same functional relationship  
reported by Corbel et al. (2003) for the BHC GX 339--4, that is
 $S_{radio}\propto (S_{X})^{0.7}$.} 
\end{figure}

Our aim was to compile (quasi--)simultaneous radio and X--ray
observations
of BHCs during the low/hard state. To this purpose, 
we have collected all the available (to our 
knowlege) data from the      
literature and made use of our own simultaneous observations as well.
These were    
taken with the Ryle telescope at 15 GHz (see 
Pooley \& Fender 1997 for more details) and combined with one--day averages
from RXTE ASM (this refers to Cygnus X--1, Cygnus X--3 and
GRS 1915+105).     
Table~1 lists the information for the ten low/hard state BHCs for which we have
at our disposal (quasi--)simultaneous radio and X--ray coverage (see
Section~\ref{beyond}, Table~\ref{tabtrans} for `non--canonical' hard state
sources, such as Cygnus X--3 and GRS 1915+105):      
distance, mass, orbital inclination, measured hydrogen column densities (see Section 2.1) and 
literature references. \\  
Both the X--ray and the radio intensity values come from several different
instruments and telescopes. Radio flux densities have been measured in
different 
frequency bands, ranging from 4.9 up to 15 GHz; nevertheless we generically
refer   
to `radio flux densities' based on the evidence that, while in the low/hard
state, black hole radio spectra are characterised by almost flat spectra ($\alpha\sim
0$) spectral index (Fender 2001a). \\    
X--ray fluxes, taken either from spectral fits or from light curves, have been
converted into Crab units in order to be   
easly compared with radio flux density units (1 Crab $\simeq$ 1060 $\mu$Jy;
energy range 2--11 keV). For this purpose, X--ray fluxes/luminosities in a
given range have been first converted into corresponding values between 2--11
keV, and then expressed as flux density.   
For those sources whose X--ray flux
has been derived from count rates, the conversion into Crab has been  
performed according to the factors provided by Brocksopp, Bandyopadhyay \&
Fender (2003).   
\subsection{Absorption corrections}
Whenever X--ray flux density has been evaluated from count rates or absorbed
fluxes, we wanted to  
compensate for absorption by calculating the ratio between the predicted flux
from a hard state BH with a measured $N_{H}$ value, and the predicted flux
corresponding to no     
absorption, as follows.    
We have first simulated with XSPEC typical spectra of hard state BHCs as
observed by \emph{Chandra} ACIS for ten diffent values of hydrogen column
density ranging from zero up to 12.5 $\times 10^{22}$cm$^{-2}$. A `typical'
spectrally hard BH's spectrum is well fitted by an absorbed power law with
photon index 1.5. 
By keeping fixed the flux corresponding to no absorption, 
the points turn out to be 
well fitted by a simple exponential relation, which allows to express the
ratio $F_{abs}/F_{unabs}$ as follows: 
\begin{equation}
\frac{F_{abs,LS}}{F_{unabs}} =  exp \Big [\frac{- (N_{H}/10^{22}\rm cm^{-2})}{18.38}
\Big ] 
\end{equation}
The procedure described has been applied to X--ray
fluxes below the transition luminosity between hard and soft state (see
Section 3.2). Above that value, the spectrum is not reproduced by a simple  
power law. In this regime, the  
X--ray spectrum is usually well fitted by an 
absorbed power law with photon index $\Gamma\simeq 2.4$ plus a disc blackbody component, with a
typical temperature of around 1 keV. Since in this case the 2--11 keV spectrum 
is almost entirely dominated by   
thermal emission, the previous simulations have been repeated for
soft state BHCs by approximating the spectrum with a disc blackbody emission
at 1 keV. We have obtained:     
\begin{equation}  
\frac{F_{abs,HS}}{F_{unabs}} =  exp  \Big [\frac{ - (N_{H}/10^{22}\rm cm^{-2})}{8.67} \Big ]
\end{equation}
The latter correction has been applied to detections above the
hard--to--soft state transition.\\   
\section{Radio vs. X--ray flux densities}
\subsection{GX 339--4 and V 404 Cygni}  
In Figure~\ref{fit} we plot radio against X--ray flux
densities  
(mJy vs.Crab), scaled to a distance of 1 kpc and absorption corrected, for GX 339--4
and V 404 Cygni, the two sources for which we have at our disposal the widest
coverage in terms of X--ray luminosity.
\\
\begin{figure*},
\centering{\epsfig{figure=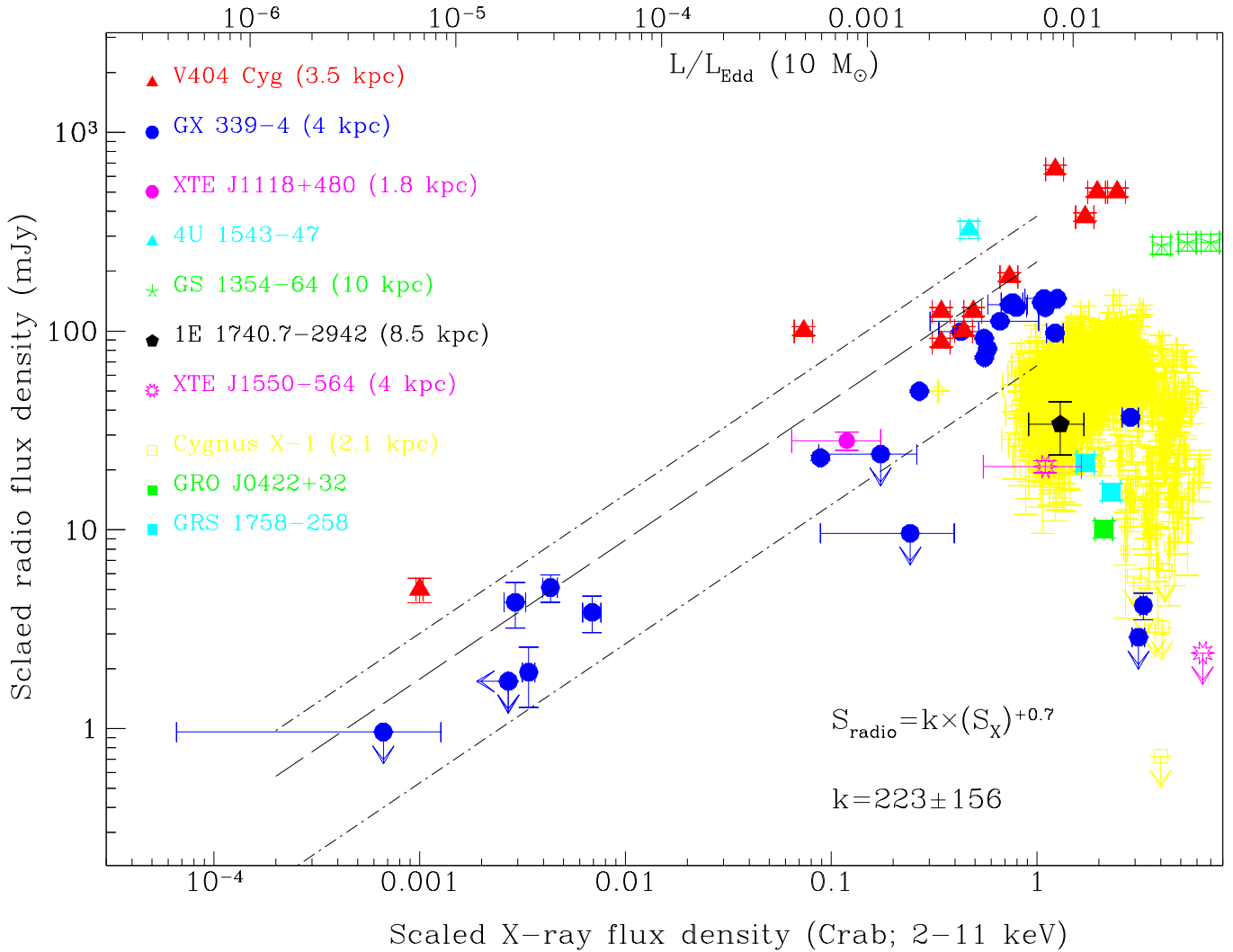,height=18cm,angle=-90}}  
\caption{\label{10bhs}Radio flux density (mJy) is plotted against X--ray flux
density (Crab) for a sample of ten hard state BHs (see Table~1), scaled to a
distance of 1 kpc and absorption corrected (this means that the axes are
proportional to \emph{luminosities}). On the top horizontal axis we
indicate luminosity, in Eddington units for a 10\msun ~BH, corresponding to the
underlying X--ray flux density. An evident correlation       
between these two bands appears and holds over more than three orders of
magnitude in luminosity. The dashed line indicates the best--fit 
to the correlation, that is $S_{radio}=k \times (S_{X})^{~+0.7}$, with
$k=223\pm 156$ (obtained by fixing the slope at $+ 0.7$, as found individually
for both GX 339--4 and V 404 Cygni; see Section
4.1). Errors are given at 3--$\sigma$   
confidence level, and arrows also represent 3--$\sigma$ upper limits.}
\end{figure*}
GX 339--4 was discovered as a radio source by Sood \& Campbell--Wilson
(1994).
When in the low/hard state, it is characterised by a flat or slightly 
inverted ($\alpha \gsim 0$) radio spectrum (see Corbel \etal
2000) and its synchrotron power has been shown to
correlate with soft and hard X--ray fluxes (Hannikainen \etal 1998; 
Corbel \etal 2000). 
By means of simultaneous radio:X--ray obsevations of GX 339--4, Corbel \etal
(2003) have recently found extremely interesting   
correlations between these two bands: in particular, $S_{8.6GHz}\propto
S_{3-9keV}^{~+0.71\pm 0.01}$  
(slightly different slopes -- within the hard state -- have been found
depending on the X--ray energy interval). When fitted in mJy vs. Crab
(scaled to 1 kpc and absorption corrected), the relation displays the form: 
\begin{eqnarray}
S_{radio}=k_{GX 339-4}\times (S_{X})^{~+0.71\pm 0.01} \\
\nonumber
k_{GX 339-4}=126\pm 3  \;\;\;\;\;\;\;\;\;\; \;\;
\end{eqnarray}
The correlation appears to hold over a
period of three years -- 1997 and between 1999--2000 -- during which the source
remained almost constantly in a spectral hard state (with a transition to the
high/soft state, Belloni \etal 1999, when the radio emission declined  
below detectable levels).   
Figure~\ref{fit} shows radio against X--ray flux densities of GX 339--4  
corresponding to simultaneous ATCA/RXTE observations performed between 1997 and
2000 (Corbel \etal 2000, 2003). Note that points above 1 Crab   
(scaled), which all correspond to RXTE--ASM detections, clearly show a 
sharp decreasing in the radio power (see next Section).  
The correlation reported by Corbel \etal (2003) 
actually refers to RXTE--PCA data only; it is worth mentioning that, 
when ASM detections below 1 Crab (\ie below the radio quenching) are
fitted together with PCA points, the final result is consistent, within the 
errors, with the fit reported by Corbel on PCA data alone (that is, a slope
of 0.70$\pm$0.06 is obtained in this case). \\ 
\\  
Remarkably, we have found that detections of V 404 Cygni,   
the source for which we have at our disposal the  
widest radio:X--ray coverage, 
are well fitted by the same functional relationship -- albeit with no
apparent cutoff -- as GX 339--4 (see   
Figure~\ref{fit}). \\ 
V 404 Cygni belongs to the class of X--ray transients, sources 
undergoing brief episodic outbursts
during which their luminosity can increase by 
a factor $\sim 10^{6}$ compared to periods of relative quiescence.
All V 404 Cygni data -- except for the quiescent lowest point -- 
come from simultaneous radio (VLA; Han \& Hjellming 
1992) and X--ray (\emph{Ginga}; Kitamoto \etal 1990)
observations during the decay following 
its May 1989 outburst, during which the
source, despite very high and apparently saturated luminosity, never entered 
a spectral soft state and always maintained a very hard X--ray spectrum
(Zycki, Done \& Smith 1999). 
According to Hjellming \etal (2000), the quiescent state of V 404 Cygni, since it
ended the long decay after its 1989 outburst, has been associated with a 0.4
mJy radio source\footnote{Starting in early 1999, VLA obervations showed
fluctuations ranging from 0.1 to 0.8 mJy on time scales of days; even more
extreme radio fluctuations in February 2000 were accompained by strong
variability in the X--ray band as well (Hjellming \etal 2000).}.    
Quiescent X--ray flux refers to a 1992  
measurement (Wagner \etal 1994 report 0.024$\pm$0.001 count/sec with
ROSAT--PSPC), \ie well before the onset of significant X--ray variability
(see Kong \etal 2002 for details).\\ 
Denoting 
S$_{radio}$ as the radio flux density in mJy and $S_{X}$  
as the X--ray flux density in Crab, we have obtained:  
\begin{eqnarray} 
\label{correq}
S_{radio}=k_{V404} \times 
(S_{X})^{~+0.70 \pm 0.20} \\
\nonumber
k_{V404}=301  \pm 43 \;\;\;\;\;\;\;\;\;\;
\end{eqnarray}   
The Spearman's rank correlation coefficient is 0.91; the two sided significance
of its deviation from zero equals 4.2 $\times 10^{-3}$.\\  \\ 
These results 
indicate that $S_{radio}\propto (S_{X})^{\sim 0.7}$ is a fundamental
property of the radio:X--ray coupling in the hard state, rather than a
peculiarity of GX 339--4. It is worth stressing that the fitted slopes
for V 404 Cygni and GX 339--4 are identical within the errors, with the same
normalisations within a factor 2.5, while detections from other sources below
1 Crab (scaled), 
although much narrower luminosity ranges, are all consistent with the same
placing in the radio:X--ray plane, as discussed in the next Section. 
\subsection{Broad properties of the correlation}
In Figure~\ref{10bhs} we plot radio flux densities (mJy) against X--ray fluxes
(Crab), scaled to a distance of 1 kpc and absorption corrected, for all the 
ten hard state BHCs listed in Table~1. Note that this scaling and correction
means that the axes are proportional to \emph{luminosities}. \\ 
Besides GX 339--4, Cygnus X--1 displays a positive $S_{radio}:S_X$
correlation followed by a radio turnover around 3 Crab, whereas 
three other sources, namely XTE  
J1118+480, 4U 1543--47 and GS 1354--64, lie very close to the relations inferred
for  GX 339--4 and V 404 Cygni. The remaining four systems (GRS 1758--258, GRO
J0422+32, 1E 1740.7--2942 and XTE J1550--564) seem instead to have already
undergone the radio quenching.\\
We can assert that these ten BH candidates display very similar 
behaviour in the  $S_{radio}$ vs. $S_X$ plane.  
There is evidence for a positive radio:X--ray correlation over more than
three orders of magnitude in terms of Eddington luminosity, as indicated in the
top horizontal axis, where we show the $L/ L_{Edd}$ ratio corresponding
to the underlying X--ray flux ($ L_{Edd}\simeq 1.3 \times 
10^{39}$~erg/sec for a 10 \msun BH).  
As an example, the total luminosity
of Cygnus X--1 in the 0.1 to 200 keV band, while the source is
in the low/hard state, is  $\sim 2\%$ of the Eddington luminosity for a 10 \msun
~BH (di Salvo \etal 2001); that corresponds to mean flux of about 2.5 Crab 
(scaled to 1 kpc).  

\subsection{Jet suppression in the soft state}
\begin{figure}
\hspace{-0.7cm}
\psfig{figure=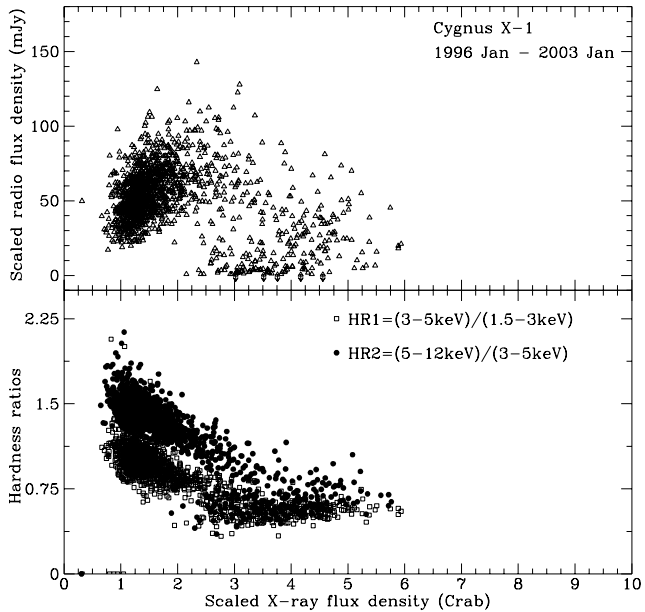,height=9cm} 
\caption{\label{hratio} Top panel: Ryle telescope at 15 GHz and RXTE ASM  
daily averages for Cygnus X-1 between January 1996 and January 2003. Fluxes have
been scaled to a distance of 1 kpc. Bottom panel: open squares for HR
(Hardness ratio)1, defined as the ratio between the count rates at 3--5 keV
and 1.5--3 keV. Filled circles for HR2, defined as the ratio between 5--12 keV
and 3--5 keV. A softening of the X--ray spectrum is shown to correspond to a 
quenching in the radio emission.} 
\end{figure}
As already mentioned, approaching the soft state radio 
emission from   
BHCs seems to be quenched below detectable levels. Such a behaviour is well
recognizable in at least two sources of our sample. As visible in Figure
~\ref{10bhs}, both Cygnus X--1 and GX 339--4 
display a clear turnover after   
which the radio power dramatically declines and reaches undetectable
levels within a factor of two in X--ray flux (see also Fender \etal 1999b). In
addition, XTE J1550--564, 
although suffering from poor statistics and large uncertainty in distance
estimates (2.5--6 kpc), 
has been detected during the hard state (Corbel 
\etal 2001; Tomsick \etal 2001) at `half--quenched' level (about 1 Crab:20
mJy scaled) while the radio emission dropped down significantly 
in the intermediate/very high state (about 6 Crab:$<$0.15 mJy
scaled). \\ 
The values of the X--ray flux density (scaled) corresponding to the turnover in radio
flux vary between the sources: GX 339--4, about 1 Crab; Cygnus X--1,
about 3 Crab; V 404 Cygni $>$ 1 Crab. 
This difference could be related to differences between the
parameters that govern the powering/quenching mechanism(s)  
of the radio emitting jet. A discriminant parameter might be the BH
mass, which is estimated to be $\sim$10 \msun ~in the case of Cygnus X--1 while it
has been recently constrained around 6  \msun ~in the case 
of GX 339--4 (Hynes \etal 2003). If this 
hypothesis is correct, we would expect not to see jet 
quenching at X--ray luminosities below the Cygnus X--1 turnover for those systems
whose BH mass has been estimated to be bigger than 10 \msun, as, for instance,
the case of V 404 Cygni, which does not appear to be quenched in radio up to
about 3 Crab, and is known to possess a 10--14 \msun BH (Orosz 2002, Hjellming \etal 2000).
Despite the poor statistics, this picture seems to
bear out the hypothesis that the X--ray luminosity at the radio quenching 
might positively correlate with the BH mass, being consistent with the jet suppression 
occurring at a constant fraction -- a few percent -- of the Eddington rate.
This is explored further in Gallo, Migliari \& Fender (in preparation).  
\\     
In Figure~\ref{hratio} the quenching of the radio power in Cygnus X--1 
(monitored simultaneously in radio and X--ray between January 1996 and January
2003) as a function of the X--ray flux density is shown together with  
X--ray hardness ratios.  
The 3--12 keV X--ray spectrum softens until about 5 Crab (scaled), whereas
radio quenching begins around 3 Crab (scaled).
While it is clear that the
quenching occurs somewhere near the point of transition from low/hard to softer
(intermediate or high/soft) states, pointed observations will be required to
see exactly what is happening to the X--ray spectrum \emph{at} the quenching point.\\
\section{Spread to the correlation}
\subsection{Best--fitting}
So far we have established two main points by looking at the distribution 
of low/hard state BH binaries in the radio vs. X--ray flux density plane:
\begin{itemize}  
\item Independently of the physical interpretation, $S_{radio} \propto (S_{X})
^{+0.7}$  
for GX 339--4 and V 404 Cygni from quiescence up to close to the
hard--to--soft 
state transition. All other hard state BHCs lie very close to these 
correlations, with similar normalisations.  
\item At a luminosity of a few percent of the Eddington rate, close to the
hard--to--soft 
state transition, a sharp turnover is observed in the radio:X--ray
relation, that is, the radio flux density drops below detectable levels. 
\end{itemize}   
Bearing this in mind, our purpose is now trying to find a
reliable expression for a `best--fit' relationship to all hard state BHs and
to estimate the spread relative to such a relation.\\ 
Assuming 0.7 as a universal slope during the low/hard state, 
we have determined the normalisation factor by fitting all the data -- Cygnus
X--1 excluded (see comments at the end of this Section) -- 
up to 1 Crab,  below which quenching does not occur for any system.\\
In this way we are able to provide an empirical relationship valid for all the
hard state BHs, that we will call   
`best--fit' in the following. We have obtained:
\begin{equation}
S_{radio}=k \times (S_{X})^{+0.7}, ~~~~~~~~~ k =223\pm 156
\end{equation}
The best--fit and its spread are indicated in
Figure~\ref{10bhs} in dashed and dot--dashed lines 
respectively. 
A scatter of about one order of magnitude in radio power 
is particularly interesting, especially in view of comparing the observed
spread to the one we expect based on beaming effects (see next
Section).\\   
\\ 
The choice of excluding the whole dataset of Cygnus X--1 
is related to its unusual behaviour in the 
radio:X--ray flux plane. In fact, a visual inspection of Figure~\ref{10bhs}
already suggested that the points below the radio power quenching belong to a
line with a steeper slope than 0.7. In addition, despite its relatively low
inclination to the line of sight (see next Section for clarity), Cygnus X--1
lies on the lower side of the correlation. It is possible that these
characteristics can be explained in terms of strong wind absorption.   
It has been demonstrated that the wind from the donor OB star in Cygnus X--1
partially absorbs the radio 
emission, up to about 10$\%$ (Pooley, Fender \& Brocksopp, 1999; Brocksopp,
Fender \& Pooley, 2002). In addition, since the jet bulk velocity during
hard state is likely to be relatively low (see Section 4.3), approaching and
receding jets contribute a similar amount to the total radio luminosity.
However, in Cygnus X--1 only a one--sided jet has been detected (Stirling
\etal 2001); this means that, possibly, a significant fraction of the 
receding jet is lost through wind absorption (however, since the jet structure
is about 100 times bigger than the orbit, it is unlikely that the wind could
absorb the entire power of the receding component). 
Furthermore, because the flat--spectrum radio emission (corresponding to steady
jet and generally associated 
with hard X--ray spectrum) is optically thick, this 
implies that when the jet power decreases, its size might also decreases 
linearly with flux. As the jet in Cygnus X--1 becomes smaller, 
the $S_{radio}$:$S_{X}$ relation will be subject to increasingly strong
wind absorption, with the net result of steepening the correlation.\\
\subsection{Constraining the Doppler factors}
What kind of physical information can we deduce from the observed relation?\\
Let us assume a very simple model, in which the same physics and   
jet/corona coupling hold for all hard state BHs and the observed functional
relationship is intrinsic; then,    
one would predict the following   
placing in the $S_{radio}$ vs. $S_{X}$ parameter space: 
\begin{enumerate}  
\item All sources lying on a line with the same slope if neither X--ray or
radio emission were significantly beamed (\ie low Doppler factors).  
\item Different sources lying on lines with the same slope but different
normalisations    
if X--rays were isotropic while radio was beamed, with radio--brighter
sources corresponding to higher Doppler factors.
\item As point (ii) but with higher Doppler factors sources being 
brighter in both radio and X--ray if both were beamed.
\end{enumerate} 
Despite the relatively small sample, we are able to place some
constraints on these possibilities. 
In the first two scenarios, where   
X--rays are isotropically emitted while radio power is beamed, we can express
the observed radio luminosity as the product of 
the intrinsic (rest--frame) radio power times the effective Doppler
factor $\Delta_{radio}$, defined as a 
function of approaching and receding 
Doppler factors\footnote{$\Delta_{radio}:=[(\delta_{app})^{2} +
(\delta_{rec})^{2}]/2$, where:\\ $\delta_{rec/app}=\Gamma^{-1}\times (1\pm \beta
\cos\theta)^{-1}$; \\ $\beta =v/c$, is the bulk velocity
of the radio--emitting material;\\ $\Gamma =(1-\beta^{2})^{-0.5}$;
$\theta$ is the inclination respect to
the line of sight.}.
If $S_{radio,intr}= k \times (S_{X})^{+0.7}$ for all hard
state BHCs, we can write:    
\begin{equation}
\label{assu}
S_{radio, obs}=\Delta_{radio}\times
S_{radio,intr}=\Delta_{radio}\times N
\times (S_{X})^{~+0.7}
\end{equation} 
Assuming the same coupling for all sources -- that means same 
normalisation N -- the ratio     
$S_{radio,1}/S_{radio,2}$ between the observed radio powers from source 1 and
2, at a fixed X--ray luminosity, will correspond to the ratio between their
relative effective Doppler factors.\\ 
\\
Returning to the case of V 404 Cygni and GX 339--4, where 
$S_{radio,V 404 }/S_{radio,GX 339-4}\sim 2.5$, we are drawn to the 
conclusion that GX    
339--4, whose inclination is poorly constrained 
between $15 < i <$ 60\degree (Cowley \etal  
2002), is likely to be located at a higher inclination than V 404 Cyg, 
well established to lie at $56\pm 4$\degree (Shahbaz \etal 1994). In fact,
assuming that $\Gamma_{GX339-4}\simeq \Gamma_{V404}$, a ratio  $S_{radio,V
404 }/S_{radio,GX 339-4}> 1$ can be achieved only if $i_{GX339-4}>i_{V
404}$.\\ 
\\
Clearly, the previous  
arguments are based on the assumption that the binary system
inclination to the line of sight coincides with the inclination of the jet,
while recent findings (Maccarone 2002) show that this is
not always the case (the misalignment  
of the disc and the jet has been already observed in Galactic jet sources GRO
J1655--40 and SAX J1819--2525). \\ 
A recent
work by Hynes \etal (2003) shows dynamical evidence for GX 339--4
being a binary system hosting a BH with mass $5.8\pm 0.5$
\msun. Interestingly, based on the spectroscopic analysis by Cowley \etal 
(2002), see their Figure 8, this value is also    
consistent with a high inclination of the system to the line of sight.   
\subsubsection{Monte Carlo simulations I: radio beaming}  
\begin{figure}
\hspace{-0.7cm}
\centering{\psfig{figure=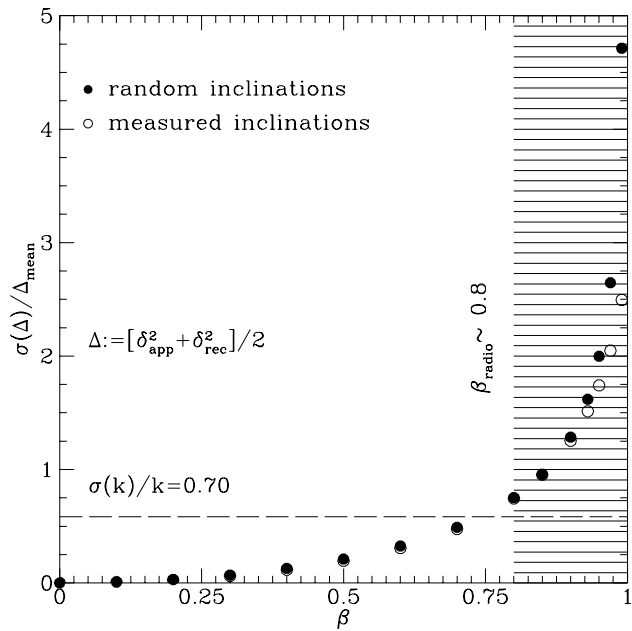,height=9cm}} 
\caption{\label{doppler}
	Assuming a simple model in which radio emission is beamed while
  	X--rays are isotropic, equation~\ref{assu} implies that  the spread in
	radio power to the best fit can be due to the distribution in Doppler
	factors. 
	The simulated spread in $\Delta_{radio}$ (defined as the ratio of 
	standard deviation over the mean value), is plotted for 15 values of
	the jet bulk velocity between 0 and $0.998c$, and compared to the measured scatter in the
	radio:X--ray correlation. In order to      
	maintain the spread in Doppler factors equal/smaller than observed,
	the outflow velocity must be smaller than 0.8 times the speed of the
	light, that is $\Gamma_{radio}<2$. Open 
	circles represent Monte Carlo simulation run by allowing the inclination
	angles to vary within the observed ranges; filled circles are for
	random inclinations between 0--90\degree. }
\end{figure} 
In order to link the measured scatter  
in the radio:X--ray relation to the beaming effects, and possibly constraining
the Doppler factors, we have performed 
Monte Carlo simulations according to the following scheme.    
\begin{table}
\label{tabmc}
\caption{\label{tabmc} 
Observed sources for which mean Doppler factors have been simulated for 10
different values of the jet bulk velocity; measured range of inclinations and
number of simultaneous radio:X--ray detections are listed. 
}
\centering
\begin{tabular}{rcl}
\hline
\hline
Source           & Inclination         & Number of      \\
                 & (degree)            &detections ($n_{j}$)\\
\hline
V 404 Cygni      &  56$\pm$4$^{a}$     &     7            \\
XTE J1118+480    &  81$\pm$2$^{a}$     &     1$^{*}$      \\
4U 1543--47      &  20.7$\pm$1.5$^{a}$ &     1            \\ 
GX 339--4        &  15--60$^{b}$       &     14           \\
\hline
\hline
\end{tabular}
\flushleft
$^{*}$ This single point corresponds to the average of 33 nearly simultaneous
detections over a very narrow interval in X--ray luminosity.\\ 
{\bf References:}  
\bf a: \rm Orosz 2002;
\bf b: \rm Cowley \etal 2002. 
\end{table}
We have considered the four sources (see Table~\ref{tabmc}) whose 
radio:X--ray flux densities have been utilised to obtain the best--fit 
relationship below 1 Crab; for each of them we have at  
our disposal a number $n_{j}$ of simultaneous detections, with a total number
of points 
$n=\Sigma_{j=1}^{4}n_{j}=23$. For each source $j$, a random inclination 
within the measured range has been generated and associated to an array whose
dimension equals the number of detections $n_{j}$, for a total of 23 random
inclinations. 
Then, for 15 different       
values of the outflow bulk velocity $\beta$, the 
corresponding  
Doppler factors $\Delta_{\beta}(j)$ have been calculated by running the
simulation $10^{4}$ times (\ie for total of $23\times 10^{4}$ values for each
$\beta$) .   
After estimating the mean Doppler factor and its relative
standard         
deviations ($\sigma(\Delta)$) for each value of $\beta$, we have compared the
simulated spread (defined     
as the ratio $\sigma(\Delta)/\Delta_{mean}$ to the measured spread in
normalisation, that is $\sigma(k)/k=156/223=0.70$. \\ 
\\ 
The result is
shown in Figure~\ref{doppler}: in order to keep the spread in Doppler factor 
equal/smaller \footnote{ 
Smaller than measured spread are also `allowed' on the ground that
errors in the distance estimates are likely to influence the observed
distribution, causing an additional source of scatter to the relation, 
which is of course not related to any boosting effect.}  
than observed (actually $\leq 0.77=0.70+10\%$), the bulk velocity of the   
radio--emitting material must be lower than $0.8 c$, that is the Lorentz
factor must be smaller than 1.7.\\ 
\\
These remarks are of course valid under the basic assumptions
that no beaming is affecting the X--rays.  
In addition we are considering a simple model in
which both the bulk velocities and opening angles of the jets are
constant (only under the latter assumption the probability of observing a
source with a given inclination $\theta$ is uniformly distributed in
cos~$\theta$). \\ 
\\
For comparison, we ran the simulation allowing the inclination angles to vary 
between 0--90\degree (filled circles in Figure~\ref{doppler}); this is actually 
crucial in the light of what has been  
discussed by Maccarone (2002) about jet--disc misalignment in BH binaries, and
also takes into account possible model--dependent errors in the estimation of
$i$. The simulated spread in Doppler factors starts to  
significantly deviate
from the previous one (calculated allowing the inclination angles for each
source to vary within the measured values; open circles in
Figure~\ref{doppler}) only for very high bulk 
velocities. As expected, even if inclinations  
as 0 and 90\degree (\ie those inclinations  
which translate into the highest and lowest possible values of cosine,
respectively) are also taken into account,
this strongly influences the mean Doppler factor only for really high values of 
$\beta$.     
\subsubsection{Monte Carlo simulations II: adding beamed X--rays} 
\label{2beam}
\begin{figure}
\hspace{-0.7cm}
\centering{\psfig{figure=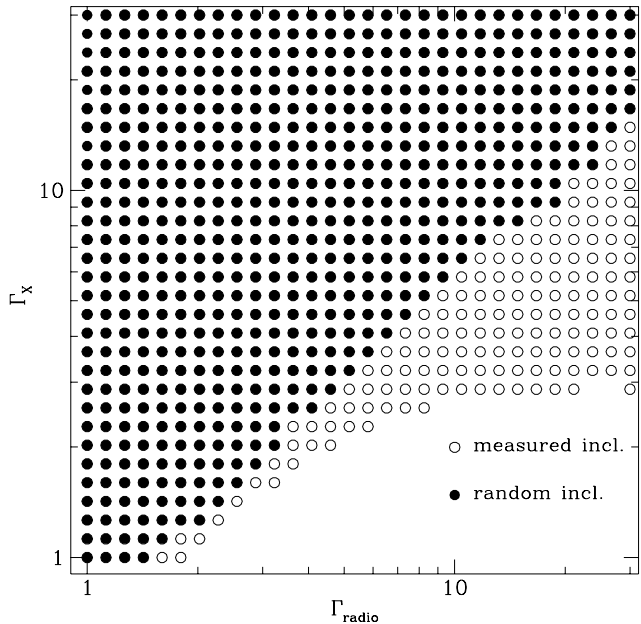,height=9cm}}
\caption{\label{doppler2} Same scheme as Figure~\ref{doppler}, but assuming a
model in which both radio and X--ray beaming are allowed; in this case, Monte
Carlo simulations have been performed with logarithmic steps  
in Lorentz factors (rather than linear in $\beta$).
The combination of radio plus X--ray beaming has the
effect of broadening the range of allowed $\Gamma_{radio}$ with respect to the case
of isotropic X--ray emission. In this picture open circles indicate 
combinations of $\Gamma_{radio}$ and $\Gamma_{X}$ for which the simulated 
spread with inclinations varying within the measured ranges is smaller than/equal
to 0.77.  
Filled circles, instead, correspond to `allowed' combinations of
$\Gamma_{radio}$ and $\Gamma_X$ when the 
inclination to the line of sight is randomly chosen between 0--90\degree.
In this case the range of possible combination is smaller due to the fact that
extreme inclinations, such as 0 or 90\degree are also taken into account.  
}
\end{figure}
Following a similar approach, it is possible to
constrain the Doppler factor due to the combination of beamed radio \emph{and}
X--ray radiation.   
As before, we assume 0.7 as an intrinsic slope which relates radio and
X--ray emission. Supposing that both X--rays and radio are beamed, we will
write: 
\begin{equation}
\label{eq_double_doppl}
\frac{S_{radio,obs}}{(S_{X,obs})^{0.7}}\propto\frac{\Delta_{radio}}{(\Delta_{X})^{0.7}}
\end{equation} 
Note that the effective Doppler factor for the X--ray radiation will be
defined as 
$\Delta_X=[(\delta_{rec})^{2.5}+(\delta_{app})^{2.5}]/2$ (assuming continuous ejection
and photon index 1.5 for low/hard state; see Mirabel \&   
Rodr\'\i guez 1994, equations 8, 9). \\
Therefore, Monte Carlo simulations have been 
run by varying both radio and X--ray bulk velocities with logarithmic steps 
in Lorentz factors, $\Gamma_{X}$ and $\Gamma_{radio}$, between 1 and 30. \\
The mean values of $ \Delta_{radio}/(\Delta_{X})^{0.7}$ together with
their standard deviations and spreads have been calculated for any combination of
the two factors. The results are shown in Figure~\ref{doppler2}.\\ 
Filled bold circles indicate those combinations of $\Gamma_{radio}$ and $\Gamma_{X}$
for which the simulated spread with inclinations varying between 0--90\degree 
is smaller than/equal to 0.77.     
Filled PLUS open circles, instead, correspond to `allowed' combinations when the 
inclination to the line of sight varies within the measured ranges.  
\\ 
In order to see why the beaming of both wavebands
allows a wider spread of $\Gamma$, consider the effect
on the position of a source in the flux--flux diagram:
a Doppler--boosted, or deboosted, radio source moves parallel to
the $S_{radio}$ axis, while a source boosted or deboosted in both wavebands
moves roughly (not exactly) parallel to the line $S_{radio} = const 
\times S_{X}$, 
and therefore roughly along the direction of the observed correlation.
Consequently beaming in both wavebands disturbs the correlation less. 
However, if X--rays were really highly beamed, that
would imply strong X--ray selection effects in detecting BH binaries.  \\
Clearly, a possible independent estimation of 
the bulk velocity of the X--ray emitting material (\eg Beloborodov 1999; Maccarone
2003) would naturally allow a much narrower constraint on the Doppler factor of
the radio--emitting material (see discussion for further details).    
\section{Beyond the hard--to--soft state transition}
\label{beyond}
\begin{figure*}
\label{transients}
\hspace{1cm}
\centering{\psfig{figure=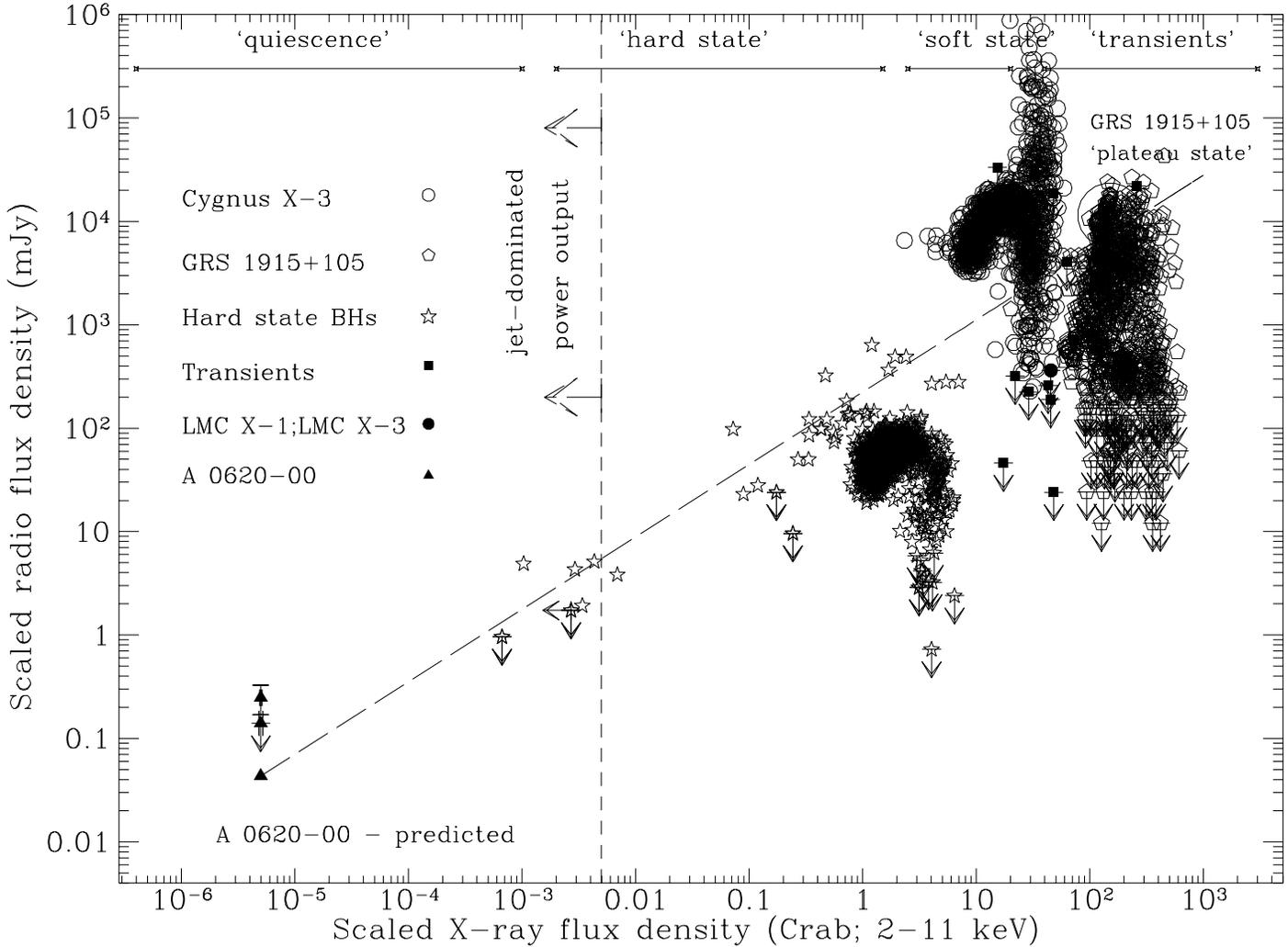,height=15cm,angle=-90}}
\caption{\label{transients}In addition to hard state BHs (open stars)   
we plot both single simultaneous radio:X--ray fluxes from black hole
transients and detections from two canonical soft state  
plus somewhat `extreme' sources, close to the Eddington accretion regime. 
Filled squares correspond to single outbursts from different sources 
(from Fender \& Kuulkers 2001); filled circles are for persistent soft/high
state sources LMC X--1 and LMC X--3,
upper limits;  
open polygons refer to GRS 1915+105, while open circles denote Cygnus X--3
points.\newline Based on the 0.7 correlation, and assuming that the radio
luminosity scales as the total jet power raised to $x=1.4$ (as in the
Blandford \& K\"onigl (1979) and MFF models), the radio luminosity is expected
to dominate 
the X--rays below $L_X\simeq    
4\times 10^{-5}\rm L_{Edd}$ (about 0.005 Crab). 
Possible detection of the nearby SXT A 0620--00 at
the predicted radio level would be capable to probe such a statement at very
low X--ray level, requiring a wholly new accretion regime on to stellar BHs.  }
\end{figure*} 
\begin{table}
\label{tabtrans}
\caption{\label{tabtrans}
Transient (plus 4 peculiar) sources whose simultaneous radio:X--ray
peak   
fluxes associated with  
discrete ejections are plotted in Figure~\ref{transients} (data from Fender \&
Kuulkers 2001).    
}
\centering
\begin{tabular}{rcl}
\hline
\hline
Source          & Distance  &       $N_{H}$ \\
                & (kpc)     &  (10$^{22}cm^{-2}$)\\
\hline
1A 0620--00     &  1   (1) &       0.2 (a)\\
GS 2000+25      &  2   (2) &       0.2 (b)\\
GS 1124--68     &  4.9 (3) &       0.5 (c) \\
GRO J1655--40   &  3   (4) &       0.7 (d) \\ 
GRS 1716--249   &  2.4 (5) &       0.4 (e) \\
GRS 1739--278   &  8.5 (6) &       2.6 (f) \\
4U 1630--472    &  10  (7) &       9.4 (g) \\
XTE J2012+381   &  10  (8) &       1.3 (h)\\
XTE J1748--288  &  8   (9) &       7.5 (i) \\
XTE J1859+226   &  7.6 (10)&       0.3 (l) \\
XTE J0421+560   &  5   (11)&       0.1 (m)\\
\hline
GRS 1915+105    &  11  (12)&       7.0 (n)\\
Cygnus X--3     &   9  (13)&       2.7 (o)\\
\hline
LMC X--1        &  55  (14)&       0.8 (14)\\
LMC X--3        &  55  (14)&       0.1 (14)\\ 
\hline
\hline
\end{tabular}
\flushleft
{\bf References :} 
\bf 1:  \rm Shahbaz \etal 1994;
\bf 2: \rm Callanan \etal 1996;
\bf 3: \rm Shahbaz \etal 1997;
\bf 4: \rm Kubota \etal 2001;
\bf 5: \rm Della Valle \etal 1994;
\bf 6: \rm Mart\'\i ~\etal 1997 ;
\bf 7: \rm Parmar \etal 1986;
\bf 8: \rm Campana \etal 2002;
\bf 9: \rm Kotani \etal 2000;
\bf 10: \rm Hynes \etal 2002;
\bf 11: \rm Robinson \etal 2002;
\bf 12: \rm Fender \etal 1999a;
\bf 13: \rm Predehl \etal 2000;
\bf 14: \rm Haardt \etal 2001.\\ 
\flushleft
\bf a: \rm Kong \etal 2002;
\bf b: \rm Rutledge \etal 1999;
\bf c: \rm Ebisawa \etal 1994;
\bf d: \rm Ueda \etal 1998;
\bf e: \rm Tanaka 1993
\bf f: \rm Greiner \etal 1996;
\bf g: \rm Tomsick \& Kaaret 2000;
\bf h: \rm Hynes \etal 1999;
\bf i: \rm Miller \etal 2001;
\bf l: \rm Markwardt \etal 1999;
\bf m: \rm Parmar \etal 2000;
\bf n: \rm Klein--Wolt \etal 2002;
\bf o: \rm Terasawa \& Nakamura 1995.
\end{table}
\subsection{Discrete ejections}
So far, we only have focused on low/hard state BHs, that is, 
binary systems characterised by a quasi--steady state of stable
accretion and whose X--ray spectrum is dominated by a hard power law. \\
In the
following we will add to our sample radio and X--ray fluxes from transient
BHCs during their episodic outbursts 
associated with discrete ejection events corresponding to optically thin
radio emission. \\  
All the available data, comprising of soft X--ray and radio
peak fluxes with references, have been reported by Fender \&  
Kuulkers (2001).  
Two points we include in this Section come 
from systems which  have also been shown
in Figure~\ref{10bhs} while in the hard state,  
namely V 404 Cygni and GRO J0422+32. 
Sources under consideration, as well as their fundamental physical
parameters, are listed in Table~3. Simultaneous radio:X--ray peak
fluxes,        
scaled to a distance of 1 kpc are plotted in Figure~\ref{transients} with
filled squares.\\
Radio data are based on peak observed flux density at frequency of 5 
GHz. Where measurements at 5 GHz were not available, a spectral index of
$\alpha = -0.5$ was assumed in order to estimate the 5
GHz flux based on observations performed at different frequencies (see Fender
\& Kuulkers 2001 for details).\\  
Most of the X--ray data are ASM detections, with the only exceptions of  GRO
J1655--40 and GRO J0422+32, whose outbursts have been detected by either
BATSE or GRANAT. 
For clarity, no error bars are plotted.
\subsection{Other BHCs: persistent soft state and `extreme' sources}
In between the jet quenching and the discrete ejections from transient
sources, there are of course binary systems displaying a persistent
soft spectrum, \ie 
whose emission is dominated by disc blackbody photons. LMC X--1 and LMC
X--3, in the Large Magellanic Clouds, 
are the only BHCs always observed while the soft state (actually Wilms \etal
2001; Boyd \& Smale 2000 and Homan \etal 2000 reported signs that LMC
X--3 entered   
in a hard state). For both these sources, the presence of a black hole is
quite well established, with a most likely mass of 9\msun ~and 6\msun,
respectively for LMC X--3 and LMC X--1. By means of radio and X--ray
observations performed in 1997, we can place the two sources 
in the $S_{radio}$ vs. $S_{X}$ plane, in order to verify the amount of radio
power from soft state BHs, which is expected to be well below the hard state 
correlation. LMC X--3 X--ray flux densities have been derived from Haardt \etal
(2001), while LMC X--1 from Gierlinski \etal 2001.
Radio upper limits for both sources are taken from Fender, Southwell \& 
Tzioumis (1998).   
The corresponding values, scaled to a distance of 1 kpc and corrected for
absorption according to equation 2, are plotted in 
Figure~\ref{transients} with filled circles (at about 150
Crab:$<$4540 mJy and   
45 Crab:$<$360 mJy, LMC X--1 and LMC X--3 respectively). As expected, both
points lie below the hard state relation extended to such high X--ray
energies. Although we are only reporting upper limits, we can assert that
fluxes from LMC X--1 and LMC 
X--3 do not disagree with our previous finding.\\ 
\\  
For completeness, simultaneous radio:X--ray detections of two
extreme sources, namely GRS 1915+105 (open polygons) and Cygnus X--3
(open circles), have been included, corresponding to the two big `clouds' at
the top right region 
of Figure~\ref{transients}. Both systems   
are traditionally considered `exotic' due to their timing
and spectral behaviours which do not fully resemble any `standard' picture
generally accetpted for BHCs; for instance, both these sources display either 
optically thin or thick radio spectra. For extensive reviews, see      
Bonnet--Bidaud \& Chardin (1988; Cygnus X--3) and Belloni \etal (2000; GRS
1915+105). Here we note that, despite its unusually high luminosity, 
detections of GRS 1915+105 in the so called \emph{plateau state} (Belloni \etal
2000) -- which appears to share similar properties with the canonical low/hard
state -- still 
seem to belong to the 0.7 correlation extended up to super--Eddington
regime.\\
\\
Cygnus X--3 is the strongest observed persistent radio--emitting BH
binary and is embedded in a    
dense stellar wind from the companion Wolf--Rayet star (van Kerkwijk \etal 
1992; Fender, Hanson \& Pooley 1999), which makes difficult
to isolate the compact object high energy spectrum. The high energy emission
from the vicinity of the compact object in Cygnus X--3 is likely to be hidden
by a dense stellar environment surrounding the source; as a consequence, the
intrinsic X--ray luminosity might be higher than inferred, pushing the dataset
closer to the 0.7 relation. Moreover, there still remains uncertainty about
the nature of the accretor in this system. 
The neutron star hypothesis can not be ruled out with confidence.\\
It is
interesting to note that, while the jets from GRS 1915+105 are at 
60--70\degree (Fender \etal 1999a), those of Cygnus X--3 appear to be close to
the line of sight 
($\simlt$14$^{\circ}$, Mioduszewski \etal 2001), supporting the previous
hypothesis, in  
which higher--than--average normalisation factors would correspond to higher
Doppler factors. In addition, the behaviour of Cygnus X--3, with the apparent
turnover in the radio power,   
is very similar to that of Cygnus X--1, except for the flaring behaviour 
following the jet quenching (note that points corresponding to this flaring
show characteristic optically thin radio spectra; on the contrary, 
pre--flaring detections display `flat' radio spectra, \ie a different physical
origin).    
\section{Predicting radio fluxes at low quiescent luminosities}    

\begin{table*}
\label{tabquie}
\caption{Five X--ray transients have been monitored in X--rays by different telescopes in
different energy ranges (indicated below the table) during quiescence. Here we
report the 
values for the maximum and the minimum luminosity, with relative references
and energy ranges.  
For these values the corresponding   
predicted radio flux densities have been calculated based upon the
radio:X--ray correlation we have 
found, that is $S_{radio}=[223\times (S_{X,1kpc} / Crab){~0.7}]/(D/kpc)^{2}$mJy.} 
\centering
\begin{tabular}{rcccl}
\hline
\hline
Source         & X--ray luminosity                     	& Distance     	&	Flux density 	&  Predicted radio\\
               & ($10^{32}$ erg/sec)                   	&  (kpc)	& at 1 kpc (10$^{-6}$Crab)& flux density ($\mu$Jy) \\
\hline 
1A 0620--00   	&0.02$^{\bf 1}$-- 0.04$^{\bf 2}$  (a,b) &   1		&1--5			&18--43\\
GRO J1655--40 	&0.2$^{\bf 3}$-- 3$^{\bf 4}$  (a,c)     &   3		&6--82			&6--34\\
XTE J1550--564	&$< 5$ $^{\bf 5}$  (d)   		&   4 		& $<173$		& $< 32$\\
GRO J0422+32  	& 0.08$^{\bf 4}$  (e)                   &   2.4        	&$\sim$ 2		& $\sim$4  \\
GS 2000+25    	& 0.02$^{\bf 4}$  (e)                   &   2          	&$\sim$ 0.5		& $\sim$2  \\
\hline
\hline
\end{tabular}
\flushleft
{\bf References :} 
\bf 1:  \rm 0.4--2.4 keV;
\bf 2:  \rm 0.4--1.4 keV;
\bf 3:  \rm 0.3--7 keV; 
\bf 4:  \rm 0.5--10 keV; 
\bf 5:  \rm 0.5--7 keV.\\
\bf a: \rm Kong \etal 2002; 
\bf b: \rm Narayan \etal 1996; 
\bf c: \rm Asai \etal 1998; 
\bf d: \rm Tomsick \etal 2001;
\bf e: \rm Garcia \etal 2001.\\
\end{table*}
As the same correlation appears to be maintained over many years and for
different 
sources (like for instance Cygnus X--1, GX 339--4, V 404 Cyg), we can 
estimate the level of radio emission from a hard state    
BH by measuring its X--ray flux alone. This is particularly 
interesting in the case of black hole X--ray transients, whose inferred
accretion rate during quiescence may be very small. \\
Kong \etal (2002) present \emph{Chandra} observations of three 
BH transients during quiescence for which no simultaneous radio detection is
available to date, namely A 0620--00, GRO J1655--40
and XTE J1550--564.   
In order to check for possible spectral variability, they also 
report results from previous X--ray observations carried out by different
telescopes, such as ROSAT, ASCA and BeppoSAX.  
According to Tomsick \etal 2001, the lowest luminosity measured
for XTE J1550--564 with Chandra ($5 \times 10^{32}$ erg/sec, for a 
distance of 4 kpc) should however be considered only as an upper limit on 
the quiescent luminosity of the system.~\emph{Chandra} detections of 
other two transient sources, namely GRO J0422+32 and GS 2000+25, are reported
by Garcia \etal 2001.  
In Table~4 we list for each of the five sources the maximum and the minimum 
measured X--ray luminosity in quiescence, the inferred distance and the predicted radio
flux density (in $\mu$Jy) based on our best--fit equation, that is
$S_{radio}=223\times (S_{X})^{+0.7}$.  
Given a spread of 156 over 223 in the normalisation factor, the predicted 
values must be considered reliable within one order of magnitude.
\subsection{A 0620--00: the ideal candidate}
The Soft X--ray Transient (SXT) A 0620--00 was discovered in outburst in 1975
August (Elvis \etal 1975), while the associated radio     
source was at a level of 200--300 mJy (Owen \etal 1976) during the onset of
the outburst.  
Six years later the source was detected with the VLA at level of
$249\pm 79\mu$Jy    
(Geldzahler 1983); additional VLA observations in 1986 
(see McClintock \etal 1995) yielded an upper limit of 140$\mu$Jy,
clearly indicating a decline in radio power (see Figure~\ref{transients}
where A 0620--00 detection/upper limit/predicted radio fluxes are marked with
triangles).       
The 1981 detection might actually be
associated to 
radio lobes resulting from    
the interaction of a relativistic--decelerating jet with the interstellar
medium, as observed in the case of XTE J1550--564 about four years after the
ejection of plasma from near the BH (Corbel \etal 2002). \\  
\\
Due to its relative proximity, A 0620--00 is the most 
suitable candidate to probe if our empirical radio:X--ray relation does hold down 
to low quiescent X--ray luminosity ($\simeq 2     
\times 10^{30}$ erg/sec at 1 kpc, \ie about 10$^{-8} L_{Edd}$; Garcia \etal
2001). In other words, if       
A 0620--00 was detected at the predicted radio level (a few tens of 
$\mu$Jy, see Figure~\ref{transients}, bottom left corner), 
it would confirm that the mechanisms at the origin of radio and X--ray emission
are correlated, if not even partly coincident, over more than six
orders of magnitude 
in X--ray luminosity. \\
\\ 
Moreover, if the radio:X--ray correlation were confirmed at very low X--ray
luminosities (below $10^{4}$ \ledd)    
it  would strongly constrain the overall theory of accretion in   
quiescence.  We direct the reader to the discussion for 
further comments.

\section{Discussion}

The presence of a coupling between radio and X--ray emission in the low/hard
state of black hole binaries obviously requires a 
theoretical interpretation that relates somehow the powering/quenching
mechanism(s) of the jet   
to the overall accretion pattern. 
Zdziarski \etal (2003) ascribe the
correlation to a correspondence between the level of X--ray
emission and the rate of ejection of radio--emitting blobs forming a compact
jet. In  
this picture, there still remains the question of the condition for jet
suppression.   
Meier (2001) interprets the
steady--jet/hard--X--ray state association as strong evidence for
magnetohydrodynamic (MHD) jet formation, where the most powerful jets
are the product of accretion flows characterised by large scale height. 
The simulations show in fact that the jet is confined by the toroidal  
component of the magnetic field lines, coiled due to the disc 
differential rotation: the bigger the disc scale--height, the stronger the
field.  
The power of the jet naturally decreases (at least 100 times weaker) 
in the soft/high state, associated with  
a standard geometrically thin disc (Shakura \& Sunyaev 1973). 
\\  
To date, two broad classes of geometries have been proposed for explaining 
X--ray emission 
from low/hard state BHs. The more classical picture is that of
a hot, homogeneous, optically thin corona of (quasi--)thermal electrons
which inverse Compton scatter `seed' photons coming from the underlying
accretion disc (Shapiro, Lightman \&  Eardley 1973 and similar later
solutions). The alternative is that of a 
jet--synchrotron model, in which, under reasonable     
assumption on the input power of the jet and the location of the first
acceleration zone, optically thin synchtrotron emission can dominate the
X--ray spectrum, traditionally thought to be a product of inverse Compton
process only (see Markoff, Falcke \& Fender 2001 for a detailed description of
the model; hereafter MFF). This model however predicts 
self--synchrotron Compton upscattering in the jet for some scenarios.
Remarkably, the
MFF model is able to reproduce the observed slope of the radio:X--ray
correlation \emph{analytically} (Markoff \etal 2003a), as a function
of the measured X--ray and radio--infrared spectral indices. \\ 
\\ 
A revised version of the classical Comptonisation model for the hard state 
has been proposed by Beloborodov (1999).  
In this case the hot coronal plasma is powered by magnetic field line
reconnections (Galeev, Rosner \& Vaiana 1979) and confined within several
active flares with mildly relativistic bulk velocities,   
inferred by the relative weakness of the reflection component. Due to
aberration effects in fact, the amount of X--rays as seen by the reflecting
disc turns out to be reduced by a factor consistent with $\beta_{X}\simeq 0.3$.\\
Maccarone (2003) draws a similar conclusion on a different ground: he has
tabulated  
all the the available measuremets of X--ray luminosities at the time of the
soft--to--hard state transition, for both BHC and neutron star systems. The 
resultant variance in state transition luminosity is also consistent with coming
from material with $\beta_{X}\simeq 0.2$; therefore $\Gamma_{X}\simeq 1$ in both
cases.\\      
Following the approach of Section~\ref{2beam}, this immediately implies a
stringent upper limit on the beaming of radio emission   
as well, that is $\bf\Gamma_{radio}\simlt 2\rm$ (see
Figure~\ref{doppler2}). This is almost certainly significantly less
relativistic than the jets produced during transients outbursts of sources
such as GRS 1915+105 (Mirabel \& Rodr\'\i guez 1994; Fender \etal 1999a) and
GRO J1655--40 (Hjellming \& Rupen 1995; Harmon \etal 1995; see also Fender
2003).\\
Therefore, if mildly relativistic beaming characterises the low/hard 
state, a mechanism(s) must exist which both switches the jet off
-- high/soft state -- and produces a faster jet -- discrete  
ejections -- above $\sim 10^{-2}$\ledd, where  
the hard--to--soft state transition occurs.  \\
\\
It is interesting to mention that 
a few percent of the Eddington rate is also close to the regime at which Ghisellini
\& Celotti (2001) have identified the transition line between FRI and FR II
radio--galaxies, the former class being associated with slower kpc--scale jets
than the latter (see 
Begelman 1982; Bicknell 1984; Laing 1993). This leads
to a kind of correspondence between `extreme sources' like GRS 1915+105, or Cyg
X--3, and FRII, characterised by quite high bulk Lorentz factors, while 
`canonical' hard state sources would be associated with FR I, with relatively
low outflow bulk velocities. \\  
\begin{figure}
\hspace{-0.7cm}
\centering{\psfig{figure=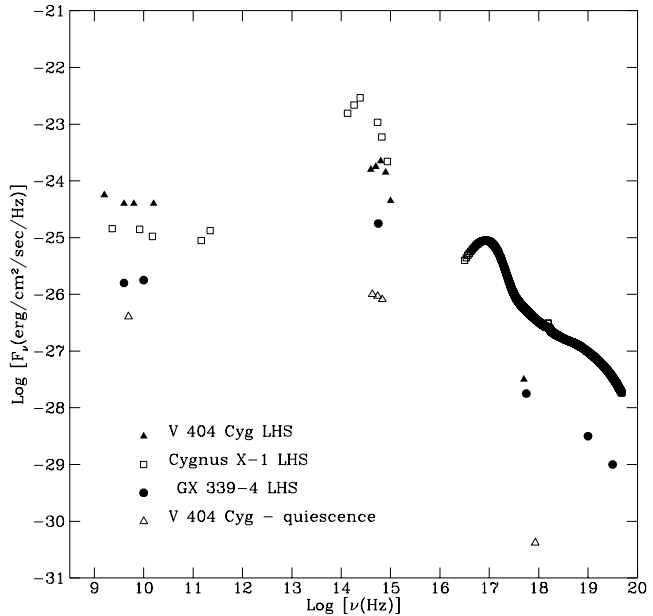,height=9cm}}
\caption{\label{broadband}Broadband spectra of Cygnus X--1, V 404 Cygni and GX
339--4: for low/hard state (LHS) black   
holes, not only the behaviours in the $S_{radio}/S_{X}$ plane look very   
similar, but even the shape of their energy spectrum, from radio wavelengths
up to X--rays, suggesting either a common origin or coupling of the basic emission
processes.     
}
\end{figure}

\subsection{Jet--dominated `quiescence'}  
A main task of the models remains that of reproducing the observational 
behaviour of accreting stellar BHs at a variety of accretion rates. An interesting
prediction of this work concerns the relative power of the jet, with respect
to the overall accretion power, at low quiescent luminosities.\\ 
In this regime, the BH spectral energy distribution appears very similar to that of
canonical hard state sources (see Figure~\ref{broadband}), although the X--ray
emission in quiescence is not well reproduced by the standard accretion--corona
model, requiring a much lower radiative efficiency.   \\ 
Narayan \etal (1996, 1997, 2001) showed that an Advection Dominated
Accretion Flow (ADAF), in which the energy released by viscous torques is
`stored' into the flow rather than radiated away, can adequately model the
available observations at  
high energies.  
However, on the other extreme of the spectrum, in the radio band, the
existence of a jet seriously weakens such a solution, requiring a significantly
different physics to model the observed spectrum.     
Jet--powered radio emission is in fact several  
orders of magnitude brighter than expected extrapolating ADAF spectra down to
radio band; moreover, the standard ADAF picture predicts highly--inverted
radio spectra, instead of the observed flat ones. \\ 
An alternative scenario for low--luminosity
stellar black holes has been proposed by Merloni \& Fabian (2002). They show that 
a coronal--outflow dominated accretion disc, in which the fraction  
of the accretion power released in the corona increases as the accretion rate
decreases, would be an ideal site for jet--launching, 
both MHD and thermally driven. \\ 
\\
Based on the radio:X--ray correlation,   
the ratio between the observed radio and X--ray luminosities 
scales as: 
\begin{equation}  
\label{-0.3}
\frac{ L_{radio}}{ L_{X}}\propto (L_{X})^{-0.3} 
\end{equation}
This already implies that, as the X--ray luminosity decreases, the
\emph{radiative} jet power will become more and more important with respect to
the X--rays. Moreover,
because of the self--absorption effects, it has been shown that there will be
not a linear relationship between the radio luminosity,   
$L_{radio}$, and the \emph{total} jet power, $L_{jet}$, for any optically thick jet
model which can explain the  
flat radio spectrum observed during the hard state. Blandford \&
K\"onigl (1979), Falcke \& Biermann (1996), and Markoff, Falcke \& Fender
(2001) obtain in fact $L_{radio}\propto (L_{jet}) ^{1.4}$. \\ 
If so, and in  
general for any relationship of the form:
\begin{equation}
L_{radio} \propto (L_{jet})^{x}     
\end{equation}
equation~\ref{-0.3} implies the following scaling for the fractional jet power:
\begin{equation}
\frac{ L_{jet} } { L_{X}}  \propto (L_{X})^{( \frac{0.7}{x} -1)} 
\end{equation}
Hence, for any $x>0.7$, there exists an X--ray luminosity below which
the jet will be the dominant output channel for the accretion power.\\ 
If $x=1.4$, by re--scaling the  
numbers to XTE J1118+480 -- emitting at 
$\sim 10^{-3} L_{Edd}$, and whose  
fractional jet power is at least 20\% (Fender et
al. 2001) -- one obtains that $L_{jet}\simgt L_X$ for $L_X \simlt
4\times 10^{-5}$\ledd 
(\ie below 0.005 Crab -- scaled -- see Figure~\ref{transients}).
Note that observations of GX 339--4 and V 404 Cygni in quiescence already
cover this regime, meaning that, if $x=1.4$, both these sources actually
\emph{are} in a jet--dominated state.\\  
This would be a wholly new      
accretion regime for X--ray binaries, requiring significant  
modification of existing (e.g. ADAF) models. In addition it would
indicate that the overwhelming majority of `known' stellar--mass black
holes, which are currently in quiescence, are in fact feeding back
most of their accretion energy into the interstellar medium in the form of the
kinetic energy of the jets and are accreting at rather higher levels than
derived based only on their X--ray luminosity.\\
If we do establish that accretion is taking place
in quiescence, for instance throught the detection of A 0620--00, 
and is furthermore channelling most of its power into 
jet formation, then the arguments for observational evidence for black
hole event horizons based upon a comparison of quiescent X--ray
luminosities of black hole and neutron star binaries (e.g. Garcia et
al. 2001) will need to be re--examined. In fact, assuming that 
$L_{jet}\propto (L_X)^{[(0.7/x)]}$ holds for both neutron stars
and black holes, then the observed difference in `radio loudness' between
black hole and neutron star binaries (Fender \& Kuulkers 2001) might be
enough on its own to explain the discrepancy, and it may be that the
event horizon plays no part. This is explored further in Fender, Gallo \& Jonker
(in preparation). 

\section{Summary }
In this paper we provide observational evidence for a broad empirical relation 
between radio and X--ray emission in Galactic black hole binaries during their
spectrally hard state. The main points established throught this work can be 
summarised as follows:
\begin{itemize}
\item In low/hard state BHCs the observed radio and X--ray fluxes are
correlated over more than three orders of magnitude in accretion rate, with a
spread in radio power of about one order of magnitude.
\item Even at accretion rates as low as $10^{-5}$ Eddington a powerful jet is
being formed; no lower limit to the relation has been found.
\item V 404 Cygni is the second source to display $S_{radio}\propto
(S_{X})^{~ 0.7}$, from quiescence up close to the hard--to--soft state transition.
\item Assuming 0.7 as a universal slope for the low/hard state, and under the
hypotheses of a) common disc--jet coupling and b) isotropic X--ray emission,
the measured spread in radio flux can be interpreted in terms of a
distribution in Doppler factors. 
Monte Carlo simulations show that the observed scatter is
consistent with relatively low beaming ($\Gamma_{radio}\simlt 2$)
outflows in the low/hard state, unlike those in transient outbursts.   
\item When the combination of radio and X--ray beaming is taken into account, 
the range of possible bulk velocities in the jet significantly broadens,
allowing the X--ray emitting material to be relativistic for almost any value of
$\Gamma_{radio}$, but implying strong X--ray selection effects. In this
case an independent estimation on $\Gamma_{X}$ is   
needed to limit $\Gamma_{radio}$. Unrelated works (Beloborodov 1999; Maccarone
2003) impose stringent constraints on the bulk velocity of the X--ray emitting
material, leading to the conclusion of relatively low radio beaming
($\Gamma_{radio}\simlt 2$) in the hard 
state. 
\item Close to the hard--to--soft state transition the jet switches
off, probably in all sources. The X--ray luminosity at
which 
the radio quenching occurs might positively correlate with the BH mass,
being consistent with taking place at a constant fraction of the
Eddington rate. It is worth mentioning that a similar fraction of \ledd~ 
has been identified as a dividing line between FR I and FR II
radio--galaxies, that is between supermassive BHs producing mildly and 
highly relativistic jets 
respectively (Ghisellini \& Celotti 2001).   
\item Since the correlation appears to be maintained over many years and for
different sources, this leads to the possibility of predicting the level of
radio emission from a hard state and/or quiescent BH by measuring its X--ray flux.    
\item If the radio luminosity scales as the total jet power raised to
$x$, with $x>0.7$, this implies the existence of an X--ray 
luminosity below which the most of
the accretion power will be channelled into the jet rather than in
the X--rays.  If  $x=1.4$ (\eg Blandford \& K\"onigl 1979, Falcke \& Biermann
1996; MFF), then below $L_X\simeq 4 \times 10^{-5}$ \ledd the jet is expected to
dominate.   
\end{itemize}
This work provides evidence for a physical coupling between radio
and hard X--ray emitting ouflows from accreting stellar BHs. 
A key, still unresolved issue concerns the modelling of the transition between
X--ray states in a self consistent way, which could possibly account for
\emph{both} the jet suppression, when the disc dominates, \emph{and} the 
transition from mildly to highly relativistic jets, as in case of transient
outbursts. Including the formation of jets   
in the overall energetics and dynamics of the accretion process at a variety
of X--ray luminosities has undoubtedly become of primary importance to
address, especially based on mounting evidence for the jet power to be a
significant fraction, if not the dominant output channel, of the total
accretion power.

\section*{Acknowledgements}
The authors wish to thank Sera Markoff, Stephane Corbel, Peter Jonker 
and Michiel van der Klis, for many useful suggestions and comments on the
manuscript. This research has made use of data obtained through the High
Energy Astrophysics Science Archive Research Center Online Service, provided
by the NASA/Goddard Space Flight Center. The Ryle telescope is supported by PPARC. 

\end{document}